\documentclass[prl,showpacs,twocolumn]{revtex4}  %% REVTeX 4.0
\usepackage{graphics}

\begin{document}

\title{Investigation and comparison of multi-state and two-state atom laser output-couplers}
\author{J. Dugu\'{e}$^{1,2}$}
\author{N. P. Robins$^1$}
\author{C. Figl$^1$}
\author{M. Jeppesen$^1$}
\author{P. Summers$^1$}
\author{J. J. Hope$^1$}
\author{J. D. Close$^1$}

\affiliation{$^1$Australian Research Council Centre Of Excellence for Quantum-Atom Optics, Physics Department, The Australian National University, Canberra, ACT 0200, Australia}
\email{nick.robins@anu.edu.au}
\homepage{http://www.acqao.org}
\affiliation{$^2$Ecole Normale Sup\'{e}rieure and Coll\`{e}ge de France, Laboratoire Kastler Brossel, 24 rue Lhomond, 75231 Paris Cedex 05, France}

\begin{abstract}
We investigate the spatial structure and temporal dynamics created in a Bose-Einstein condensate (BEC) by radio-frequency (RF) atom laser output-couplers using a one-dimensional mean-field model. We compare the behavior of a `pure' two-state atom laser to the multi-level systems demonstrated in laboratories. In particular, we investigate the peak homogeneous output flux, classical fluctuations in the beam and the onset of a bound state which shuts down the atom laser output.   
\end{abstract}


\maketitle
\section{Introduction}
The analogy between atom lasers and optical lasers is strong \cite{wise}. Both optical and atom lasers create a coherent output beam of bosons that are photons in the case of an optical laser and de Broglie matter waves in the case of an atom laser. The lasing mode in an optical laser is pumped through a non-thermal equilibrium process and is not the lowest energy mode of the cavity but, rather, is a highly excited mode many wavelengths long. For atoms, the lasing mode is populated through Bose-Einstein condensation and is the ground state of the trap. In an optical laser, a beam is outcoupled using a grating or a partially reflecting mirror. Most atom laser beams studied to date have been outcoupled either continuously (long pulse) or pulsed using RF radiation to transfer atoms from a trapped magnetic sub-state to a state that does not interact with the trapping field \cite{mewes,bloch,aspect}. The atoms  fall away from the trap under gravity producing a coherent matter wave \cite{ottl,kohl,bloch2,kasevich1}. 

Optical lasers have found broad applications in precision measurements to address questions both fundamental and applied in nature. In many cases, we expect to be able to perform such experiments more effectively and to higher precision with atom interferometry \cite{kasevich, berman}. Measurements made with the low density, highly coherent beam from an atom laser may not be limited in precision by the mean-field interaction that plagues interferometric measurements made with more dense atomic sources such as full Bose-Einstein condensates \cite{ketterle,aspect1}. 

We have recently found that the RF output-coupler has a number of properties which hinder the production of a high flux, shot-noise limited atom laser \cite{n,n1}. We have found that a trade off must be made between output flux and classical (or dynamical) fluctuations in the output beam. We have measured that the peak homogeneous flux into the atom laser beam is significantly below the flux that can be provided by the finite reservoir of BEC atoms in a typical experiment. This homogeneous flux limit is imposed by the interaction between the inherent multiple internal Zeeman states of the magnetically confined atoms. Furthermore, we have also found that a previously predicted effect known as the bound state of an atom laser \cite{jeffers} effectively shuts off the RF output-coupler and hence the atom laser beam.  

Recently, two-state atom laser systems have been produced experimentally \cite{wow, aspect} with the aim of avoiding classical fluctuations existing in more complicated multi-state systems. Y. Le Coq {\it et al.} produce a $^{87}$Rb BEC in the $F=1, m_F=-1$ state in a harmonic trap with a large bias magnetic field (40G). An atom laser is created by applying an RF coupling resonant with a transition from the $m_F=-1$ state to the $m_F=0$ state. Transitions to the anti-trapped $m_F=1$ state are suppressed by the second order Zeeman shift, creating an effective two-state atom laser. A. \"Ottl  {\it et al.} also create a $^{87}$Rb BEC in the $F=1, m_F=-1$ state. The authors then use a microwave transition at 6.8GHz to resonantly couple atoms to the $F=2, m_F=0$ state. The magnetic trap bias field splits the Zeeman states in the F=2 manifold by approximately 1MHz, once again leading to a two-state system.  

In this paper we utilize a one-dimensional numerical model to investigate the spatial dynamics of an RF output-coupler for five, three and two-level atom laser systems. The five and three-state systems correspond to the experimentally relevant Zeeman levels of the F=2 and F=1 ground states of  $^{87}$Rb and the two-state simulations to the recent experiments of Y. Le Coq {\it et al.}  and A. \"Ottl  {\it et al.}. The question we answer is whether the significant amount of experimental effort to make a two-state system has an actual influence on the quality of the atom laser. We analyze each of these systems with respect to experimentally important properties. In particular, we investigate the peak homogeneous output flux, classical fluctuations in the beam and the onset of a bound state which shuts down the atom laser output. 

We find that, in the weak coupling regime, the three-state and pure two-state system are indistinguishable with respect to classical fluctuations and flux. We conclude that the added experimental complications of producing a truly pure two-state system do not represent a significant benefit.

\section{The model}
An RF output-coupler uses the resonance between non-degenerate Zeeman states in the small bias field at the minimum of the magnetic trap. Applying a monochromatic RF magnetic field resonant with this splitting coherently couples atoms between the relevant Zeeman states. The magnetically trapped  $F=2, m_{F}=2$ atoms can then be `shunted' through the  $F=2, m_{F}=1$ state to the  $F=2, m_{F}=0$ state in which they no longer interact with the magnetic potential. The presence of the trapping magnetic field also introduces a spatial resonance associated with a given frequency allowing the centre of the resonance to be tuned within and around the condensate. The resonance condition is satisfied on the surface of an ellipsoid centered around the minimum of the magnetic field. Gravity introduces an asymmetry such that the minimum of the trapping potential given by $\frac{1}{2}m\omega^2_\rho\rho^2$ is shifted down vertically by $G_{shift}=g/\omega^2_{\rho}$ from the centre of the magnetic field minimum. Here m is the atomic mass, g the acceleration due to gravity and $\omega_\rho$ the radial trapping frequency of the $F=2, m_{F}=2$ state. This asymmetry produces a `preferred' direction of the coupling process, and the atoms fall out of the trapping region.  
\subsection{Time dependent equations}

Ballagh {\it et al.} \cite{ballagh} introduced the Gross-Pitaevksii (GP) equation as an effective tool for investigating this type of atom laser within the semi-classical mean-field approximation. A number of groups have found good agreement between theory and experiment \cite{steck,zhang, schneider}, using mean-field models of the atom laser. Numerical descriptions of an atom laser within the mean-field frame work are complicated by the large velocities that atoms reach when falling in a gravitational potential. The resultant small de Broglie wavelengths require very fine temporal and spatial numerical grids in order to accurately follow the dynamics. Additionally, to run simulations that reflect experiments on a time scale longer than a few milliseconds, an apodising mechanism must be introduced to absorb the beam and hence avoid breakdown of the numerical techniques used. These factors make simulating and understanding atom laser dynamics a complicated proposition. In order to simplify the numerics,  a true 3D system can be transformed to lower dimensions. The dimensionality reduction can be performed non rigorously by writing an equivalent equation for the system in the dimension(s) of interest \cite{schneider}. The F=2 GP model \cite{n} of the atom laser in one dimension (the coordinate over which gravity acts) is then given by 

\begin{equation}
 \label{secondset}
\begin{array}{l} 
{\displaystyle
i\dot\phi_2=({\mathcal{L}}+
z^2+Gz-2\Delta )\phi_2+2\Omega\phi_1}\\[7pt]
{\displaystyle
i\dot\phi_1=({\mathcal{L}}+
\frac{1}{2}z^2+Gz-\Delta )\phi_1+2\Omega\phi_2+\sqrt{6}\Omega\phi_0}\\[7pt]
{\displaystyle
i\dot\phi_0=({\mathcal{L}}+Gz) \phi_0+\sqrt{6}\Omega\phi_1+\sqrt{6}\Omega\phi_{-1}}\\[7pt]
{\displaystyle
i\dot\phi_{-1}=({\mathcal{L}}
-\frac{1}{2}z^2+Gz+\Delta )\phi_{-1}+2\Omega\phi_{-2}+\sqrt{6}\Omega\phi_0}\\[7pt]
{\displaystyle
i\dot\phi_{-2}=({\mathcal{L}}-
z^2+Gz+2\Delta )\phi_{-2}+2\Omega\phi_{-1} \,  ,}
\end{array}\end{equation}

where $\phi_i$ is the GP function for the $i$th Zeeman state and ${\mathcal{L}}\equiv-\frac{1}{2}\frac{\partial^2}{\partial z^2}+U (\Sigma_{i=-2}^{2} |\phi_{i} |^2)$. Here $\Delta$ and $\Omega$ are respectively the detuning of the RF field from the $B_0$ resonance, and the Rabi frequency, measured in units of the radial trapping frequency $\omega_z $ (for the $F=2, m_F=1$ state), $U$ is the interaction coefficient and $G=\frac{mg}{\hbar\omega_z }(\frac{\hbar}{m\omega_z })^{1/2}$ gravity. The wave functions, time, spatial coordinates, and interaction strengths are measured in the units of $(\hbar/m\omega_z)^{-1/4}$, $\omega_z ^{-1}$, $(\hbar/m\omega_z )^{1/2}$, and $(\hbar \omega_z )^{-1}(\hbar/m\omega_z )^{-1/2}$, respectively. The nonlinear interaction strength is derived by requiring that the 1D Thomas-Fermi chemical potential be equivalent to the 3D case. There are no free parameters in this model; we use  $U=6.6\times10^{-4}$, $G=9.24$, $\Omega=0-14$, $\Delta=10.7$. Although all the numerics are executed in dimensionless units, data are presented in dimensional units.

Fortuitously, the $g_F$ factors for the F=2 and F=1 ground states of $^{87}Rb$ are the same except for a change of sign. The three-state and two-state atom laser systems are then simply subsets of the five-state equations presented above, with the modification that the pre-factor $\sqrt{6}$ in the coupling terms between the $m_F=-1,0,1$ states becomes $\sqrt{2}$.
 
Our simulations are performed in a commercially available and widely used platform Matlab, using a fourier based, symmetric split step algorithm. For this work we use a 2048 point spatial grid from -40 to 40 with the equilibrium position of the condensate at 20 and a temporal resolution of $10^{-4}$. Apodising boundaries for each state have an exponential form and are positioned in order that no spatial aliasing occurs.  

In addition to solving the time-dependent GP equations for the atom laser we also find the time-independent solutions to provide accurate initial conditions for our simulations. The time-independent GP equation can be found from Eq.\ref{secondset} using the substitution $\phi_2({\bf r},t)=\varphi({\bf r})e^{i\mu t}$.  We use a relaxation technique \cite{numerical_recipes} which, if required, allows us to find excited stationary states as well as the ground state solution.  For the five-state system we use as our initial condition the solution to the following time-independent equation,
\begin{equation}
 \mu_2=(-\frac{1}{2}\frac{\partial^2}{\partial z^2}+U_2  |\phi_{2} |^2+z^2)\phi_{2}.
\end{equation}
For the two and three-state systems we solve 
\begin{equation}
 \mu_1=(-\frac{1}{2}\frac{\partial^2}{\partial z^2}+U_1  |\phi_{-1} |^2+\frac{1}{2}z^2)\phi_{-1}.
\end{equation}
In these equations the one-dimensional interaction strengths are calculated by requiring that the one and three-dimensional chemical potentials are equal.

\section{Results}
In this section we present the general results for each of the five, three and two-state systems.  Unless otherwise stated, for each system studied the RF coupling resonance is arranged to be at the center of the trapped BEC by selecting an appropriate value for the detuning.
\subsection{Five-state system}

We first analyze the five-state system. In figure 1 we show the population dynamics for strong and weak coupling.  Although we are using an apodising boundary in the numerical simulations we keep track of the total population in each state by calculating at each timestep the one-dimensional flux passing through a point (we like to think of it as a detector) on the numerical grid, significantly below the trapped condensate and above the apodising boundary.  Hence, the numerical grid is effectively divided into two sections: above and below this 'detector'.  At any time during the simulation we can determine the total number of atoms in a particular state by summing the number of atoms on the grid above the detector with the total number of atoms that have passed through the detector.  Apart from giving results that are more intuitively obvious, this method allows us to monitor the normalisation of the numerics. A number of theoretical works have suggested that the atom laser may have a `bound'  eigenstate \cite{hope,jeffers,moy}, based purely on the existence of coupling between a single trap mode and a continuum of un-trapped states.  Furthermore, in the context of producing a two-dimensional BEC, it has been shown recently that trapping of all $m_F$ states is a natural consequence of combining RF coupling with a DC magnetic trap \cite{zobay,co}.  This trapping can be understood by considering the `dressed-state' basis in which the RF coupling and DC potentials seen by the atoms are diagonalised.   In this basis the dressed eigenstates are linear combinations of the bare Zeeman states, trapped in effective potentials created by the avoided crossings.  Assuming the strong coupling limit and a sudden non-adiabatic projection onto the dressed-states, diagonalization yields a prediction of up to 62.5\% of the initial condensate atoms remaining trapped for the F=2 atom laser (four of the five dressed-states allow some trapping).

In figure 1(a) the condensate (trapped atoms in the $m_F=2$ and $m_F=1$ states) decays slowly and monotonically, with some small modulation introduced by the output-coupling process itself.  This behavior is indicative of the intermediate output-coupling regime in which most experiments have been operated \cite{bloch, aspect, n1}.  In contrast, in figure 1(b) we see the formation of a bound state in the case of strong output-coupling.   At $t=0$ the high power RF coupling is switched on and the $m_F=2$ Zeeman state is projected onto the new dressed-state basis as discussed above.  After a short ($\sim 1ms$) high frequency exchange a fraction of un-trapped dressed-states are ejected, leaving the remnant dressed-states to oscillate in the magnetic trap.

Although the population dynamics in figure 1(a) is smooth, we anticipated that the details of the spatial dynamics would not be. This is because in the five-state system atoms trapped in the $m_F=2$ state must pass through the $m_F=1$ state to get to the un-trapped $m_F=0$ state.  Because atoms in the $m_F=1$ state have a different equilibrium position to the $m_F=2$ state (the gravitational sag is different for each state) the $m_F=1$ atoms start to oscillate in the magnetic trap. The dynamics of such an oscillation is shown in figure 2. This phenomena is independent of the output-coupling strength, and indicates that even at low flux a five-state atom laser system will be modulated by classical noise. Density fluctuations in the output beam are shown figure 3, for different Rabi frequencies in the weak coupling regime.  We note here that the spatial dynamics imposed on the atom laser beam by the mechanism of the $m_F=1$ sloshing in the trap are an independent noise mechanism compared to the back-coupling dynamics investigated previously \cite{n}.

\subsection{Three and two-state systems}

The three-level system offers the possibility of a cleaner output than the five-level, as there is no intermediate state between the trapped condensate ($m_F=-1$ Zeeman state) and the un-trapped beam ($m_F=0$ Zeeman state).   Fluctuations in the output will then be due to the back-coupling and depletion of atoms to the anti-trapped $m_F=1$ Zeeman state as observed previously \cite{n1}.  However, in the limit that the output-coupling becomes weak these effects will be negligible and the system should produce a classically quiet atom laser beam.  In figures 4 and 5 we show the population dynamics for the weak and strong coupling regimes for the three and two-state systems respectively.  The behavior of the three and two-state population dynamics is qualitatively similar to the five-state system.   For weak coupling strengths the condensate decays monotonically, with a greater fraction of condensed atoms transferred to the $m_F=0$ state than the five-state system under similar conditions.   In the strong coupling limit the behavior of the system is  similar to the five-state, although here there is a more clean cyclic oscillation of atoms between the three Zeeman states.  Once again a percentage of all states remain trapped.  

The oscillatory population dynamics observed in figures 1, 4 and 5 (b) for the strong coupling limit is also reflected, if not driven by, periodic spatial oscillations.  In figure 6 we show an example of such oscillations for the five-state system.  In this figure one see the ejection of the un-trapped dressed-states early in the simulation, and then a clean periodic oscillation.  Two oscillation periods are clear from this figure, corresponding to the F=2 and F=1 radial trapping frequencies.  Interestingly, these oscillations have their upper maximum positions at the minimum of the magnetic trapping potential (the point of gravitational sag for the F=2 state), and hence oscillate up only one side of the total potential (trap plus gravity).   Similar behaviour occurs for the two and three-state systems as well.  We believe that further detailed theoretical and experimental study of these unusual dressed-state spinor (multicomponent) condensates will reveal a rich and potentially useful phase space.   Already, dressed systems have been used to demonstrate two-dimensional trapping potentials \cite{co} and to investigate double well interferometry with condensates \cite{chips}.  

\subsection{Comparison between two and three-state systems} 

In this section we compare the dynamics of the two and three-state atom lasers.  As mentioned in the introduction, significant effort is required in order to produce a `closed' two-state atom laser as the relevant alkali atom manifolds have at least 3 Zeeman states linked by allowed RF transitions.  Given that back-coupling fluctuations and the bound state arise even in the two-level system, limiting the homogeneous output flux as in the `natural' systems, it is prudent to ask whether the two and three-state atom lasers actually differ that much in the weak coupling limit. 

In figure 7(a) we compare the time dependent density of the two and three-state atom lasers at a point below the condensate.  It is clear that, for the one-dimensional model we are using, the two systems are essentially indistinguishable in the weak coupling regime.  It is interesting to note that even for stronger coupling,  the details of the classical fluctuations on the beam are also mirrored in the two systems.  The difference in amplitude between the simulations is accounted for by the loss of atoms to the anti-trapped Zeeman state in the three-level system.   In the weak coupling limit these atoms are expelled from the system on a time scale much faster than the back-coupling time set by the Rabi frequency, and so have little effect on the dynamics of the system.

In figure 7(b) we show the density fluctuations as a function of time for the two or three-state atom laser.  It is clear that even in the weak coupling limit taken here the density fluctuations due to back-coupling are a significant contribution to the noise in the atom laser beam.   This has major implications for the use of atom lasers in precision measurement systems, where high flux and minimal classical noise are essential features required of the atom beam. 

In figure 8 we compare the number of atoms in the condensate to the number of atoms in the $m_F=0$ atom laser state for 15 ms of output-coupling. There is a clear peak in the number of atoms transferred into the  $m_F=0$ state as a function of Rabi frequency for all systems considered in this paper, with the two-state system having the highest coupling efficiency and the five-state having the lowest.   However, a comparison of figure 8 with figure 7 reveals that we will not be able to operate the atom laser near the peak output-coupling rate because of increasingly severe density fluctuations, which impede the usefulness of the beam for measurement applications. For example, the peak in the two-state system is around $\Omega=500Hz$ while the density fluctuations are already significant at a Rabi frequency of 130Hz.

In summary, we find that the peak homogeneous output-coupling rate achievable in an atom laser to be significantly lower than the maximum output-coupling rate.  For the one-dimensional model considered here we find that, in this homogeneous output-coupling regime, there is practically no difference in our results between the three and two-state atom lasers.  One may ask whether the small amount of anti-trapped atoms generated in the three-state system will effect the results of a correlation function measurement on the atomic beam, such as the one made recently in the Zurich group \cite{wow}.  In this experiment an atom laser beam is produced in the weak coupling regime and subsequently falls through a high finesse cavity, in which the researchers are able to detect the passage of a single atom with high temporal resolution.  A simple calculation shows that the small amount of anti-trapped atoms produced by a three-state atom laser system would miss the cavity by a large margin, thus having no effect on the results of the measurement statistics for the $m_F=0$ state.  One can imagine many experiments where a true two-level system is important, but we conclude that the atom laser is not one of them, at least in the weak coupling limit.

\subsection{Discussion of results with respect to future experiments}

We plan to lead further experiments in the near future.
By creating a very large condensate we should be able to have both resonance positions (of the $m_F=2$ and $m_F=1$ trapped states) inside the condensate. One can expect the spatial dynamics imposed on the atom laser beam by the mechanism of the $m_F=1$ sloshing would then disappear. However, preliminary simulations carried in this configuration seem to show no difference and further investigation is necessary.

A Raman output-coupling setup will also be implemented as it overcomes many problems with RF output-couplers \cite{n2}. In particular, in the case of a Raman atom laser, we could selectively outcouple atoms in a given Zeeman state, taking advantage of the spatial dependance of the resonances.

\section{conclusion}

In this paper, we have characterized how RF output-coupling affects the spatial structure and temporal dynamics in a BEC. We also presented the 1D mean-field model used for this purpose. The peak homogeneous output flux, classical fluctuations in the atom laser beam and the onset of a bound state were investigated in the cases of multi-state and two-state atom laser systems. The results showed the five-state system was clearly inappropriate to create a clean and homogeneous atom laser beam. It also appeared the two and three-state systems were almost indistinguishable in the weak coupling regime where the atomic beam is homogeneous, with no density fluctuations. As a conclusion we think the `natural' three-state system should be preferred as it does not require any additional experimental devices.

We are currently studying a momentum kicked Raman coupling based on the same model and it will be interesting and important for future measurement devices to compare the results to those described in the present paper.

\section{acknowledgments}

We would like to thank S. A. Haine and M. T. Johnsson for fruitful discussions relevant to this work. This research was supported by the Australian Research Council Centre of Excellence Programme.

\begin{figure*}[]
\centerline{\scalebox{.4}{\includegraphics{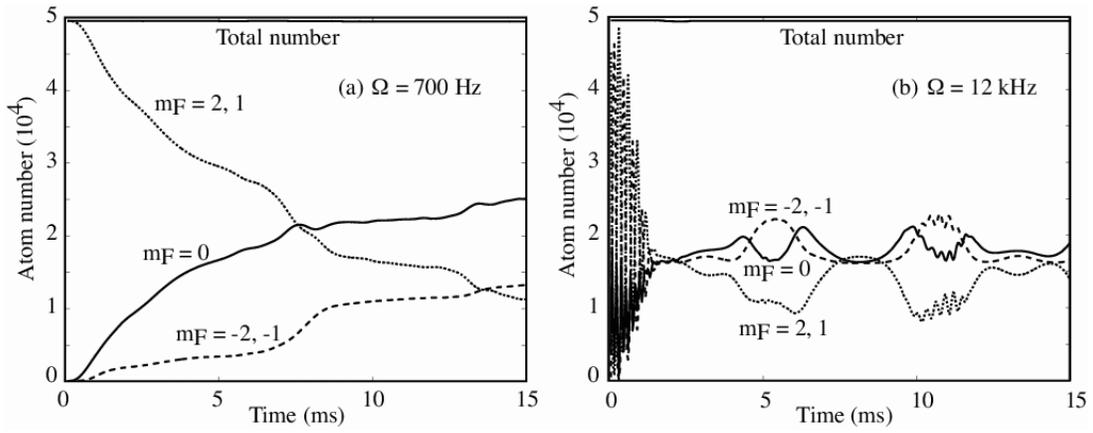}}}
\caption{Comparison of the population dynamics in the five-state system for (a) weak and (b) strong coupling. Parameters are $U=6.6\times10^{-4}$, $G=9.24$, $\Delta=10.7$.}
\end{figure*}
\begin{figure*}[]
\centerline{\scalebox{.27}{\includegraphics{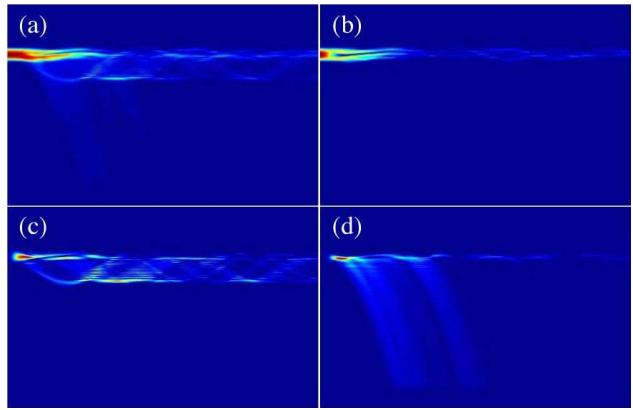}}}
\caption{Spatial profiles of the five-state system, showing the effect of the $m_F=1$ state oscillations on the output mode.  The color axes are adjusted to give the best contrast for each of the Zeeman states.  Each plot shows the atomic density, with the complete spatial grid in the vertical direction and the temporal grid in the horizontal. (a) Total density (the sum of the densities in all five Zeeman states) as a function of time, (b) density of the $m_F=2$ state, (c) density of the $m_F=1$ state, and (d) $m_F=0$ state. The dynamics were obtained after 15 ms of intermediate output-coupling ($\Omega= 1 kHz$).}
\end{figure*} 
\begin{figure*}[]
\centerline{\scalebox{0.45}{\includegraphics{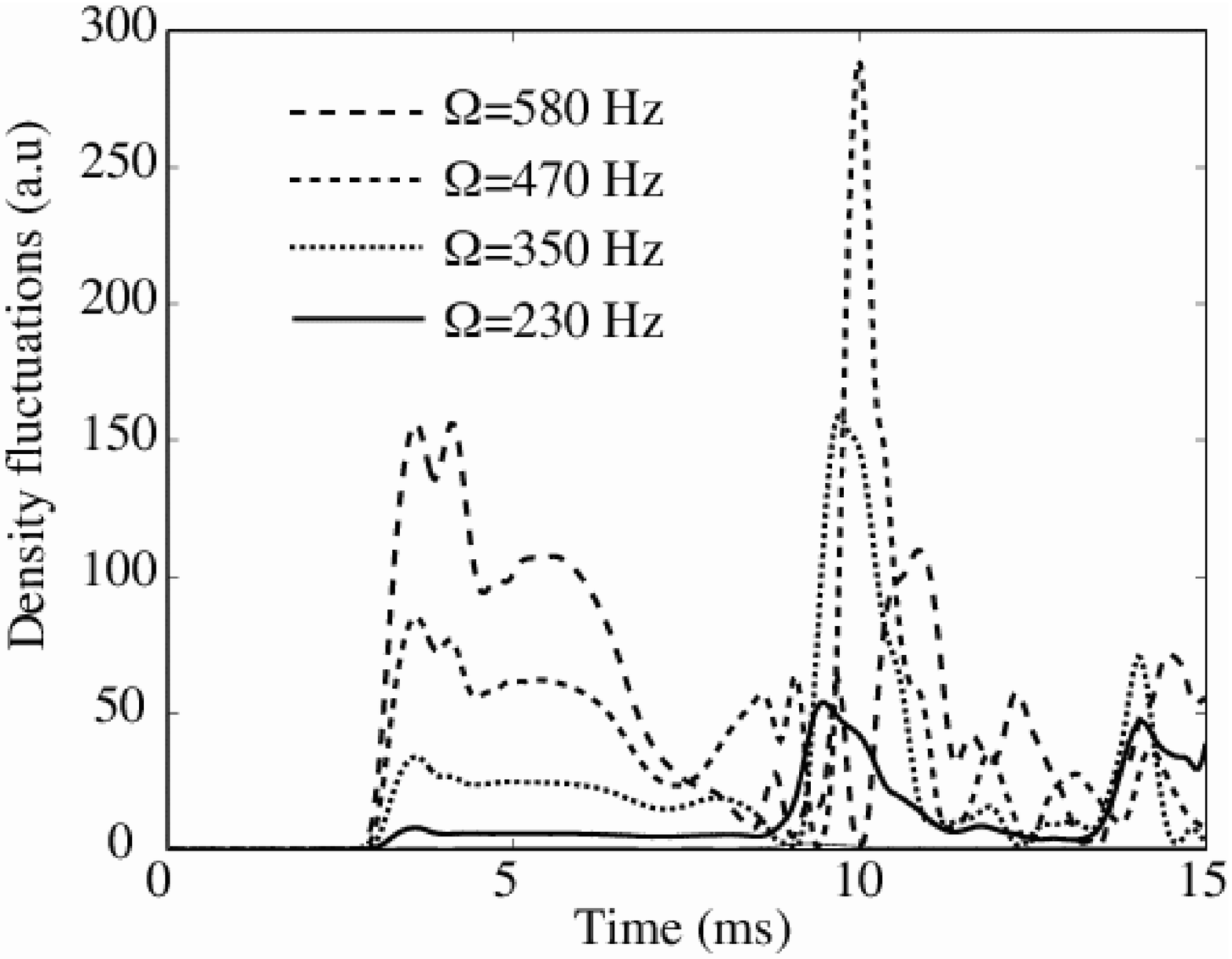}}}
\caption{Density fluctuations in the $m_F=0$ state of a five-state system at a single point in the beam as a function of time.  The fluctuations due to the sloshing of the $m_F=1$ are increasingly severe as the coupling strength is increased. Parameters are $U=6.6\times10^{-4}$, $G=9.24$, $\Delta=10.7$.}
\end{figure*}
\begin{figure*}[]
\centerline{\scalebox{.4}{\includegraphics{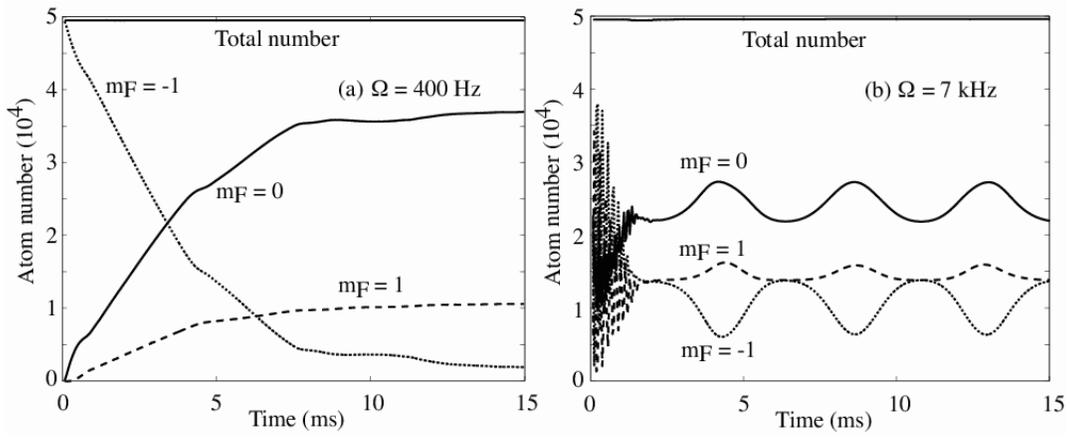}}}
\caption{Population dynamics in the three-state system for (a) weak and (b) strong coupling. Parameters are $U=6.6\times10^{-4}$, $G=9.24$, $\Delta=43$.}
\end{figure*}
\begin{figure*}[]
\centerline{\scalebox{.4}{\includegraphics{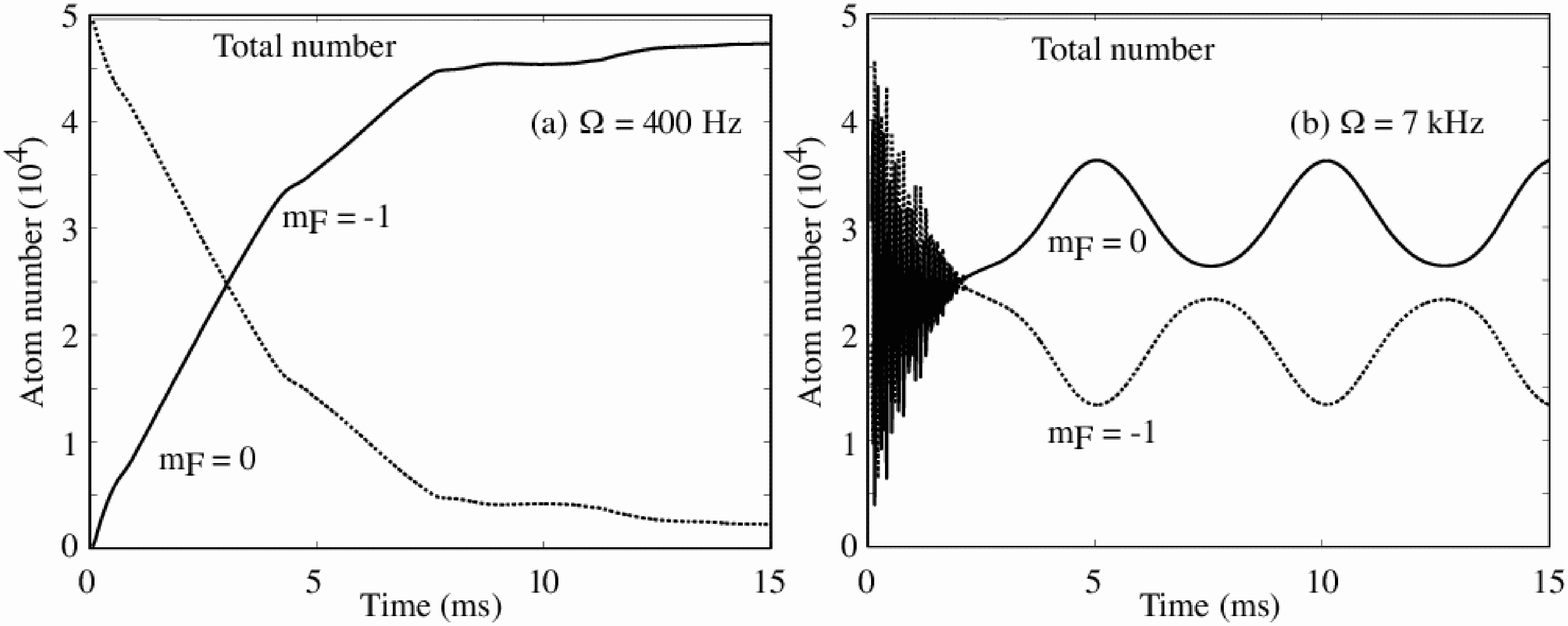}}}
\caption{Comparison of the population dynamics in the two-state system for (a) weak and (b) strong coupling. Parameters are $U=6.6\times10^{-4}$, $G=9.24$, $\Delta=43$.}
\end{figure*}
\begin{figure*}[]
\centerline{\scalebox{.7}{\includegraphics{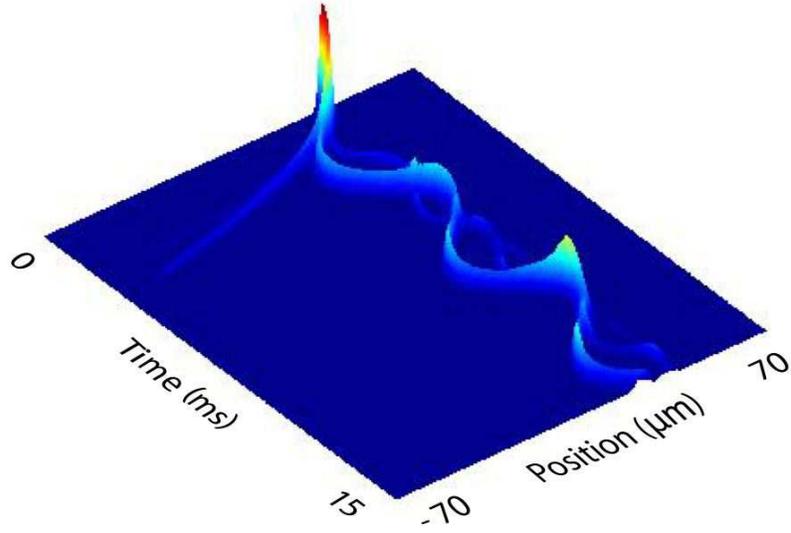}}}
\caption{Periodic spatial oscillations in the population dynamics of a five-state system. One can see the ejection of the un-trapped dressed-states as well as both oscillation periods corresponding to the F=2 and F=1 radial trapping frequencies.  Note that the upper edge of the oscillations is at the minimum of the magnetic trapping potential.}
\end{figure*}
\begin{figure*}
\centerline{\scalebox{.4}{\includegraphics{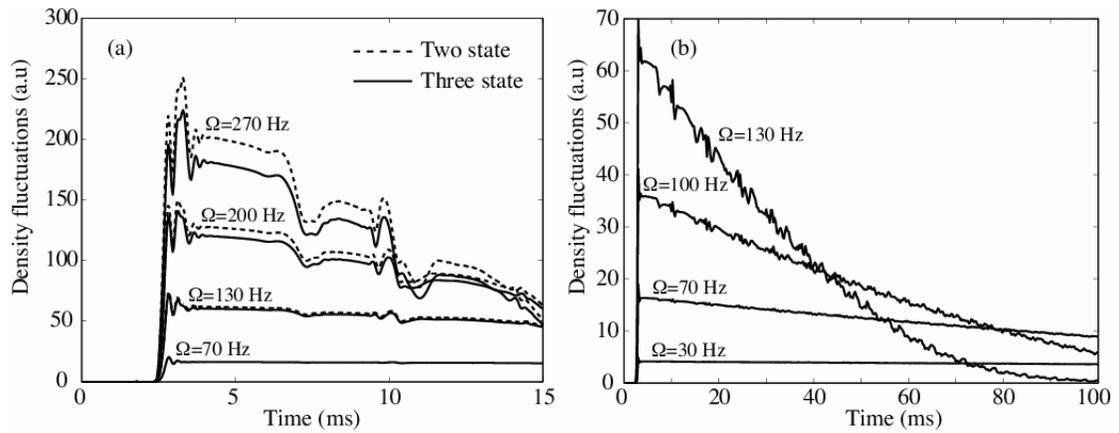}}}
\caption{Comparison of the time-dependent density of the two and three-state atom lasers at a point below the condensate after 15 ms of output-coupling and for different Rabi frequencies (a). (b) shows the time-dependent density of the two-state atom laser at very low intensities after 100 ms of output-coupling.}
\end{figure*}
\begin{figure*}
\centerline{\scalebox{.4}{\includegraphics{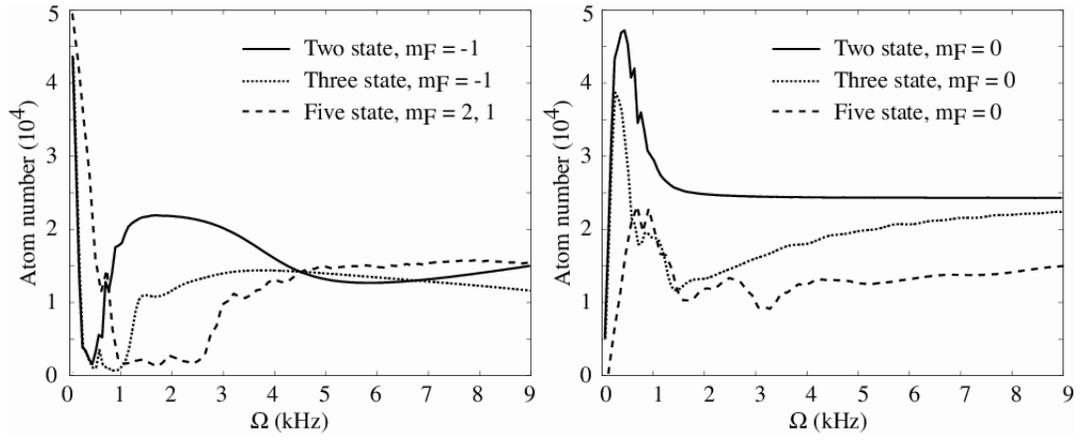}}}
\caption{Comparison of the number of atoms in the condensate and in the $m_F=0$ atom laser state for the two-state and multi-state systems after 15 ms of output-coupling.}
\end{figure*}

\end{document}